# Equally Spaced Quantum States in van der Waals Epitaxy-Grown Nano-Islands

Chaofei Liu,[†] Chunxiang Zhao,[‡] Shan Zhong,[†] Cheng Chen,[†] Zhenyu Zhang,[§] Yu Jia,[‡] and Jian Wang[†,#,¶,*]

[†]International Center for Quantum Materials, School of Physics, Peking University, Beijing 100871, China
[‡]International Laboratory for Quantum Functional Materials of Henan, and School of Physics and Microelectronics, Zhengzhou University, Zhengzhou 450001, China
[§]International Center for Quantum Design of Functional Materials (ICQD), Hefei National Laboratory for Physical Sciences at Microscale, and Synergetic Innovation Center of Quantum Information and Quantum Physics, University of Science and Technology of China, Hefei 230026, China
[#]CAS Center for Excellence in Topological Quantum Computation, University of Chinese Academy of Sciences, Beijing 100190, China
[¶]Beijing Academy of Quantum Information Sciences, Beijing 100193, China

*jianwangphysics@pku.edu.cn.

**ABSTRACT:** Pursuing the confinement of linearly dispersive relativistic fermions is of interest in both fundamental physics and potential applications. Here, we report strong STM evidence for the equally spaced, strikingly sharp, and densely distributed quantum well states (QWSs) near Fermi energy in Pb(111) nano-islands, van-der-Waals epitaxially grown on graphitized 6$H$-SiC(0001). The observations can be explained as the quantized energies of confined linearly dispersive [111] electrons, which essentially 'simulate' the out-of-plane relativistic quasiparticles. The equally spaced QWSs with an origin of confined relativistic electrons are supported by phenomenological simulations and Fabry–Pérot fittings based on the relativistic fermions. First-principles calculations further reveal that the spin–orbit coupling strengthens the relativistic nature of electrons near Fermi energy. Our finding uncovers the unique equally spaced quantum states in electronic systems beyond Landau levels, and may inspire future studies on confined relativistic quasiparticles in flourishing topological materials and applications in structurally simpler quantum cascade laser.

For atomically flat ultrathin films, the energies of the vertically confined standing-wave-like electrons are quantized into the discrete quantum well states (QWSs). With increasing the film thickness, various properties can be periodically modulated by the QWSs, such as density of states (DOS), film thermodynamic stability,[1-3] tunneling current,[4] electron–phonon coupling,[5] superconducting transition temperature $T_c$,[6] work function,[7] and surface diffusion,[8] etc. The studies of QWSs have inspired the invention of quantum cascade laser, typically achieved from the repeated emissions between the nearly parallel, low-energy QWSs subbands in multiple stacked heterostructures of semiconductor quantum wells.[9]

Essentially, such quantum cascade laser occurs based on the QWSs of 'free' electrons satisfying $E = \frac{(\hbar k)^2}{2m}$ that have been generally introduced in textbooks. By contrast, the quantum confinement of relativistic fermions with linear dispersion ($E = \hbar k v$), e.g. Dirac/Weyl particles, would be of more fundamental interest with exotic physics and can in principle yield the quantum cascade laser in a single quantum well.[10, 11] The Dirac materials can 'harbor' the relativistic fermions.[12, 13] For example, the quantum confinement of relativistic Dirac fermions has been intensively studied in the archetypical 2D Dirac semimetal, graphene.[14-22] In experiments, the signatures of such Dirac confinement are identified as the nonperiodic Coulomb blockade peaks in patterned graphene quantum dots,[14] and the in-plane spatial DOS modulations in graphene nano-islands[15-17, 19] and graphene quantum dots defined by circular PN junctions.[18, 20-22] Whereas, the QWSs of relativistic fermions are rarely investigated systematically in systems beyond graphene.

While the QWSs of 'free' electrons exhibit the gradually enlarging intervals with enhancing the energy level, the QWSs of relativistic electrons are equally spaced,[23] offering a very simple approach in differentiating the non-relativistic and relativistic fermions. Along the experimentally preferred [111] ($k_\perp$: Γ–L) direction, the band calculations for bulk Pb(111) show isolated, approximately linear dispersions $E(k_\perp)$ within a large energy window of ~[−2,4] eV.[24, 25] Thus, the Pb(111) provides an ideal system for studying the confinement of relativistic electrons beyond graphene. However, in previous studies, i) the number of QWSs probed by STM is low, mainly limited by the film thickness [≲ 30 monolayer (ML); 1 ML = 2.86 Å],[4, 26-32] and ii) the lineshapes of QWSs detected by ARPES are highly broadened with ill-identified peak positions.[5, 6, 33-36] The quantitative proof of the precisely equal interval of QWSs is still lacking as the evidence for confined linearly dispersive relativistic electrons.

Here, in the MBE-grown quasi-free-standing Pb(111) nano-islands (~60 ML), we report the spectroscopic observation of the strictly equally spaced QWSs by using *in situ* STM. The [111]-oriented Pb nano-islands[37-41] were synthesized via van der Waals epitaxy growth on bilayer graphene (BLG) on graphitized 6H-SiC(0001) (Figures 1a and 1b; see Methods). To be specific, the deposited Pb atoms nucleate and grow directly into islands in the Volmer–Weber mode, resulting in the [111]-oriented Pb nano-islands (Figure S1) as evidenced by the hexagonal-type low-energy electron-diffraction pattern.[41] Dynes fittings to the temperature-dependent spectra taken on the Pb nano-islands show the BCS-type superconducting gaps at different temperatures with $T_c$ = 6.65 K, which manifests the high quality of the Pb nano-islands (Figures 1c–1e).

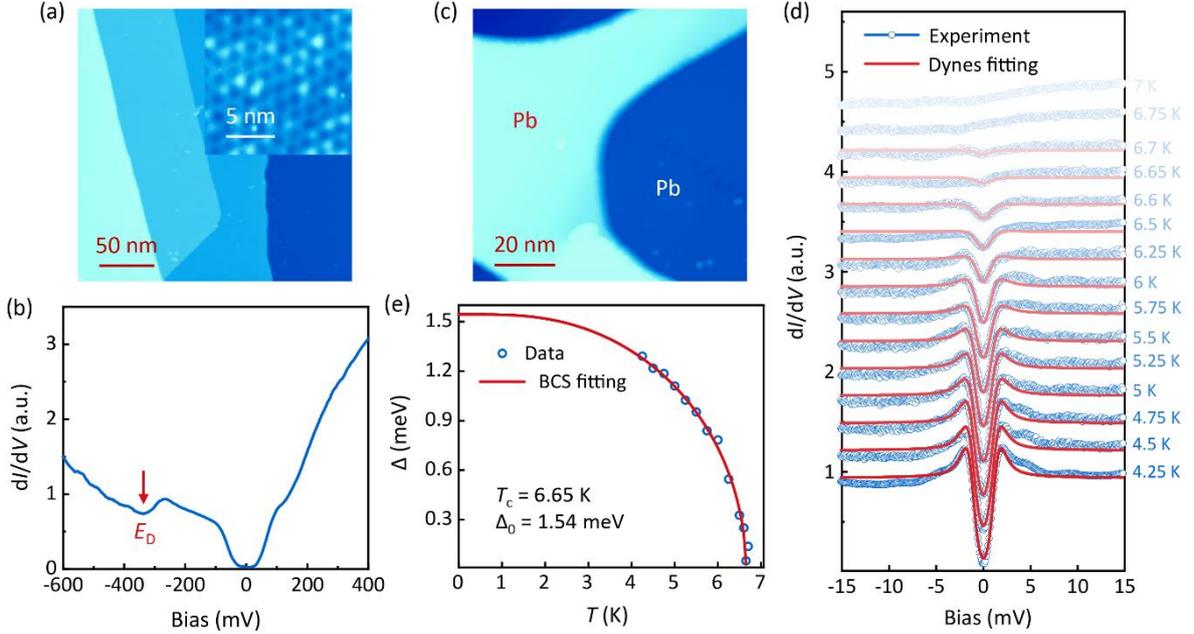

**Figure 1.** Pb nano-islands grown on BLG. (a) Topographic images of the graphitized SiC in different scales. The surface shows separated steps with typical width of ~50 nm and height of 0.75 nm. After the standard flash-annealing process,[42] the AB-stacked BLG is epitaxially formed on the graphitized SiC and periodically modulated by the interfacial C-rich 6×6 reconstruction (inset).[43, 44] Size: main, 250×250 nm$^2$, inset, 15×15 nm$^2$; set point: main, $V$ = 1 V, $I$ = 500 pA, inset, $V$ = 0.1 V, $I$ = 1500 pA. (b) Typical tunneling spectrum taken on epitaxial BLG, revealing $E_D$ near −0.33 eV. $E_D$, Dirac energy. a.u., arbitrary unit. Set point: $V$ = 0.04 V, $I$ = 500 pA; modulation: $V_{mod}$ = 5 mV (by default). (c) Topographic image of the Pb nano-islands grown on BLG (nominal coverage, 6.7 ML). Size: 90×90 nm$^2$; set point: $V$ = 0.8 V, $I$ = 500 pA. (d) Temperature dependence of the experimental tunneling spectra (open symbols; vertically offset for clarity) with superconducting lineshape, which are taken on the Pb nano-island (left part) in (c). The spectra are well fitted by the Dynes function (solid curves), $\frac{dI}{dV} \propto \int_{-\infty}^{\infty} \frac{1}{k_B T} \cosh^{-2}\frac{E+eV}{2k_B T} \text{Re}\left[\frac{|E-i\Gamma|}{\sqrt{(E-i\Gamma)^2-\Delta^2}}\right] dE$ ($\Gamma$, quasiparticle inverse lifetime; $\Delta$, superconducting gap).[45, 46] Set point: $V$ = 0.04 V, $I$ = 2500 pA; modulation: $V_{mod}$ = 0.3 mV. (e) BCS fitting (solid curve) by $\Delta(T) = \Delta_0 \tanh\left(\frac{\pi}{2}\sqrt{\frac{T_c}{T}-1}\right)$[47] to the temperature-dependent gap $\Delta(T)$ (open symbols) obtained from the Dynes fittings in (d), yielding $T_c$ = 6.65 K, and $\Delta_0$ = 1.54 meV. $\Delta_0$, $\Delta$ at 0 K. Accordingly, the BCS ratio $\frac{2\Delta_0}{k_B T_c}$ is 5.4, slightly higher than 4.4 for bulk Pb.

For all the measured Pb nano-islands, the well-defined QWSs (at 4.25 K by default) were detected (e.g., Figures 2a and 2b), essentially due to the vertical confinement of Pb-6$p_z$ electrons.[41] Preliminarily, the observed QWSs seem equally spaced (Figure 2b), which is more clearly seen in the normalized spectrum (Figure 2c). By plotting the energy positions $E_n$ of QWSs vs. their indexes $n$, the linear behavior with correlation coefficient $r$~1 is obtained (Figure 2d; $r$ =

0.9997), suggesting the strictly equal interval Δ$E$ between adjacent QWSs. The slope of linear fitting to the extracted $E_n$–$n$ curve yields Δ$E$ = 226.8 meV. As an independent proof of the periodicity of QWSs, the Fourier analysis of the normalized QWS spectrum reveals the sharp low-frequency peak (Figure 2e), highlighting the well-defined periodicity. Quantitatively, the low-frequency component gives Δ$E$ = 222.2 meV, comparable with that (226.8 meV) extracted by the linear fitting. Consistently, Fourier analyses of raw and background-*subtracted* QWS spectra both yield similar Δ$E$ with negligible changes (Supporting Information Part SII). To check the variation of tunneling spectra in space, we collected the spatially resolved line spectra within both larger ([−1,1] V) and smaller ([−15,15] mV) bias windows, both showing the spectral uniformity at nano-meter scale (Supporting Information Part SIII). The detected QWSs within [−1,1] eV here are strictly equally spaced, strikingly sharp and densely distributed, which are different from previously reported STM results,[26-32, 48] and establish the central finding of our study.

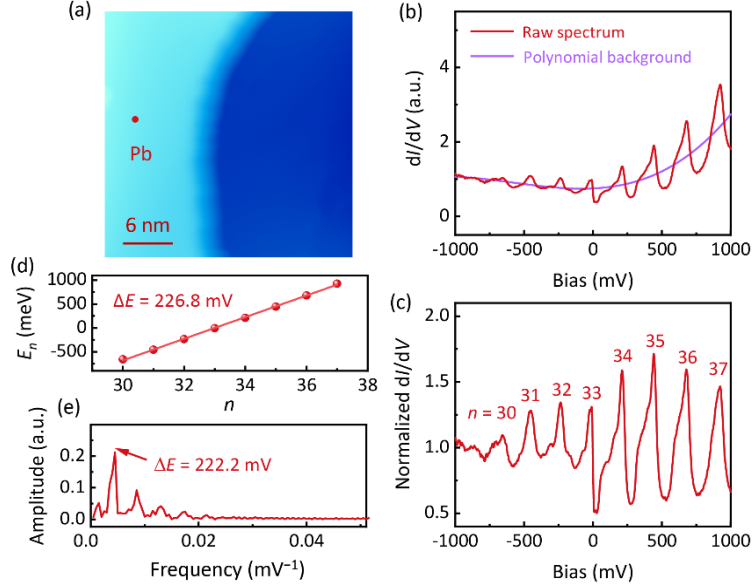

**Figure 2.** (a) STM images of the Pb nano-island (nominal coverage, 6.7 ML). Size: 30×30 nm$^2$; set point: $V$ = 0.3 V, $I$ = 500 pA. (b) Tunneling spectrum taken on the position in Pb nano-island as indicated in (a), showing the sharply defined QWSs. The light violet curve is the corresponding cubic-polynomial background, which is similar to previous works.[7, 29] (c) Normalized QWS spectrum, obtained via dividing the original spectrum by its polynomial background. The indexes $n$ of QWSs are assigned by considering that, the round-off integer [$n$] of the well width $w$ ($w$ ≈ 61 ML here; obtained as described in Supporting Information Part SV) divided by $\lambda_F$/2 ($\lambda_F$, Fermi wavelength) represents the quantum number of the highest occupied QWS just below Fermi level.[26] (d) Energy positions $E_n$ of QWSs vs. indexes $n$ (solid symbols). The slope of linear fitting (solid line) gives the energy interval Δ$E$ of QWSs. (e) Fourier analysis of the normalized QWS spectrum in (c). The low-frequency peak (arrow) also gives Δ$E$.

Statistically, besides the 'normal'-case QWS spectrum in Figure 2, there exist the QWSs modulated by the Pb nano-islands with discrete structures (Figures 3 and S4). Specifically, the modulated QWSs show the fine structures of 'satellite' peaks residing near the 'main' QWSs (Figures 3b and 3e). The energy positions $E_n$ of main QWSs linearly depend on their indexes $n$ regardless of the satellite states (Figures 3c and 3f), suggesting the main QWSs remain equally spaced. Notably, the satellite peaks (e.g. for $n$ = 36–39 QWSs in Figure 3b) evolve periodically in the tunneling spectrum (see Supporting Information Part SIV for detailed description). Such QWSs superimposed with gradually evolving satellite peaks can be understood by two sets of QWS peaks with different periods for the nano-islands showing two discrete thicknesses. (As illustrated in Figures 4b and 4c below, the satellite peaks of QWSs can be reproduced by superimposing the QWS spectra with two different periods.) The double-thickness quantum well responsible for these modulated QWSs may originate from the stacked-island structures (e.g., see Figures 3a and 3d),[23] or the wedged, top-surface-flat Pb islands[8] grown over vicinal SiC steps. In the stacked-island scenario, electrons confined within the upper island contribute to one set of QWSs. Due to the non-ideal reflection at interface between the two stacked islands, electrons partially transmitting into the lower island bounce back and forth vertically within the whole stacked structure, and contribute to the other set of QWSs with a different period. For these 'stacked' QWS spectra (i.e., Figures 3b and 3e), a specific region (dashed ellipse) can be identified, which indicates the crossover energy window for the QWS lineshape transiting between QWSs peaks with and without evident satellites. In the double-period picture, such transition region corresponds to the accidental degeneracy of QWSs for the two involved QWS spectra with different periods. The above tunability of QWSs manifests the Pb islands–BLG/SiC as a potential system for the quantum-state engineering.

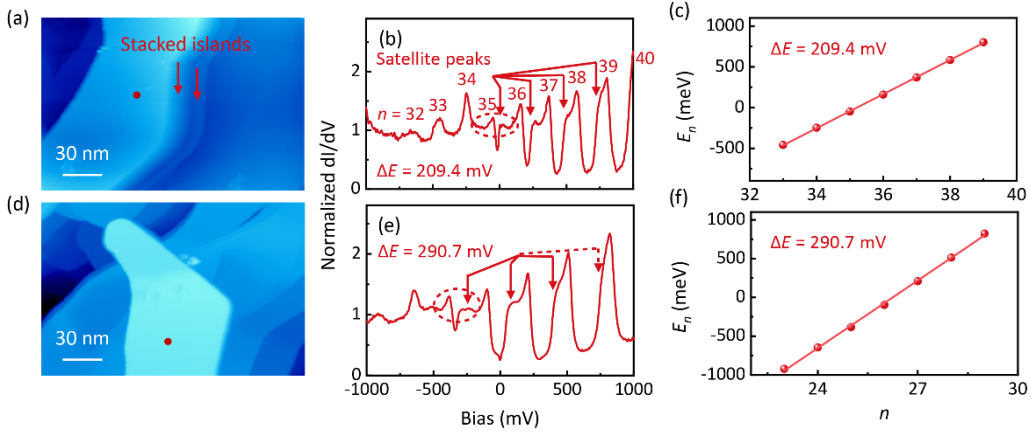

**Figure 3.** (b,e) Normalized QWS spectra taken on the positions in Pb nano-islands as indicated in (a,d) (nominal coverages, 6.7 ML), respectively. (For calculated well widths $w$ and QWS indexes [$n$], see Table S1.) The satellite peaks (arrows) and the transition regions (dashed ellipses) of QWS lineshape are highlighted. (a,d) Size: 200×132 nm$^2$; set point: $V$ = 1 V, $I$ = 500 pA. (c,f) Energy positions $E_n$ of main QWSs vs. indexes $n$ (solid symbols) for the spectra in (b,e), respectively. The energy intervals $\Delta E$ of QWSs are given by the slopes of linear fittings (solid lines) to $E_n$ vs. $n$.

To interpret the QWSs theoretically, we simulated the DOS of 'normal' and 'stacked' QWSs by a phenomenological model,[23] which is based on the Lorentzian-broadened quantized energy levels with equal interval for the relativistic fermions (see Methods). As shown in Figures 4a–4c, the simulations reproduce all types of equal-interval QWS spectra within the relativistic scenario (Figure 4a vs. Figure 2c; Figures 4b and 4c vs. Figures 3b and 3e, respectively), particularly the details including satellite peaks, transition region of QWS lineshape, and energy-dependent QWS amplitude. These results clearly reveal the equally spaced QWSs as the quantum confinement of the linearly dispersive Γ–L relativistic quasiparticles in Pb(111) nano-islands.[5, 24, 32] (For the exclusion of BLG as the origin of relativistic electrons, see Supporting Information Part SVI.)

Fabry–Pérot fittings to the QWS spectra can provide the independent evidence for the relativistic-electron-like confinement. Essentially, the electrons confined in a quantum well are analogous to the electromagnetic standing waves in an optical Fabry–Pérot interferometer filled with an absorptive medium. Following the spirit, we fitted the main peaks of the measured QWS spectra by the Fabry–Pérot formula (see Methods). From Figures 4d–4f, only for linear $k_\perp(E)$ relation (Figure 4g), the fits are basically in agreement with all types of equally spaced main QWSs, again suggesting the confinement of relativistic fermions. [For square-root $k_\perp(E)$ relation (i.e., parabolic $E(k_\perp)$), the Fabry–Pérot formula yields unequal-spacing QWSs as expected and fails to fit the QWS spectra (Part Supporting Information SVII).]

Meanwhile, the full information of $E$-dependent $k_\perp$, Γ, $R$ and Φ is firstly quantitatively determined for Pb quantum wells (Figures 4g–4j), providing crucial insights into the QWS electrons. (Here, Γ, quasiparticle inverse lifetime; $R$, product of reflectivities at two boundaries; Φ, total electron-scattering phase shift at surface and interface; see Methods.) i) $k_\perp$. The band dispersions $E(k_\perp)$ cross Fermi energy $E_F$ at the Fermi wavevector $k_F/k_{\Gamma L}$ = 0.52–0.56 (Figure 4g; $k_{\Gamma L}$, Γ–L momentum), which is very close to $k_F/k_{\Gamma L}$ = 0.55 obtained from the first-principles calculation.[24] The obtained Fermi velocity $v_F$ based on $E = \hbar k_\perp v_F$ is 1.89×10$^6$–2.03×10$^6$ m/s, again consistent with $v_F$ = 1.9×10$^6$ m/s along Pb [111] direction.[4, 49] ii) $R$. $R$ is less than unity (Figure 4i), as expected for the non-ideal confinement induced by non-specular reflection at the Pb island–BLG/SiC interface. iii) Γ. Besides the relatively negligible finite-temperature and tunneling broadenings ($\Gamma_T = \sqrt{(2.5eV_{\mathrm{mod}})^2 + (3.5k_BT)^2} = 12.6$ meV), the QWS peaks are further broadened by lossy interface reflections,[25] resulting in the observed peak width $\Delta_E$ (~0.1 eV) far larger than $\Gamma$ (Figure 4h: $\Gamma \lesssim 50$ meV). iv) $\Phi$. The phase shift $\Phi$ changes maximally by ~$\pi/2$ over $[-1,1]$ eV (Figure 4j). Such strikingly lower $\Phi$ suggests the negligible image potential[50] in Pb nano-islands (Supporting Information Part SVIII).

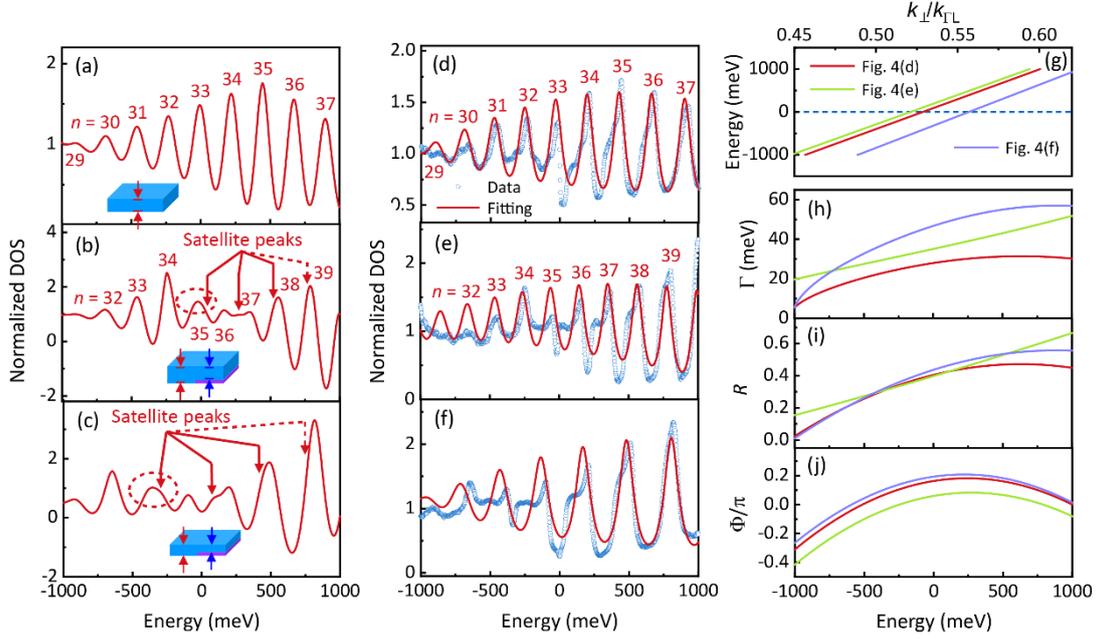

**Figure 4.** (a–c) Simulated DOS spectra of 'normal' (e.g. spectrum in Figure 2c) and 'stacked' QWSs (e.g. spectra in Figures 3b and 3e) for relativistic fermions (see Methods). The insets are the schematic well structures for QWS modeling. $\Delta E$ shown in Figure 2d and Figure 3 are adopted in the modeling. $\delta_E = 0.1$ eV; $\delta_E$, spectral peak width. (d–f) Fittings to the main QWSs of 'normal' and 'stacked' spectra by Fabry–Pérot formula (see Methods). (g–j) Obtained out-of-plane momentum $k_\perp/k_{\Gamma L}$, inverse lifetime $\Gamma$, reflectivity $R$ and phase shift $\Phi$ from Fabry–Pérot fittings.

To study the strong spin−orbit coupling (SOC) effect of Pb on QWSs, we carried out first-principles calculations of Pb(111) films with different thicknesses (see Methods). The relativistic calculations were performed using scalar-relativistic Kohn–Sham equations,[51] with and without SOC for comparison. Figure 5a presents the calculated energies of QWSs at $\Gamma$ point as a function of thickness and the corresponding $\Gamma$L dispersion. As the thickness increases by every two layers, there will be a branch (solid line) moving down and crossing the Fermi level (i.e., at $E_F$, the inter-branch interval $\approx$ 2 ML), which is in agreement with the previous discovery of bilayer stability.[52]

Interestingly, for all calculated thicknesses, the SOC strengthens the trend of QWSs near $E_F$ towards being more strictly equally spaced. Taking the 61- and 65-ML Pb(111) as examples (comparable with ~61-ML Pb nano-island in Figure 2), when SOC is included, the energies $E_n$ of QWSs vs. their indexes $n$ indicate more strict linear behavior with improved correlation coefficient $r$ closer to 1 (Figure 5b; specifically, for 61-ML Pb film: $r_{\mathrm{without\ SOC}} = 0.99881$, $r_{\mathrm{SOC}} = 0.99993$; for 65-ML Pb film: $r_{\mathrm{without\ SOC}} = 0.99989$, $r_{\mathrm{SOC}} = 0.99994$). Consistently, the inclusion of SOC suppresses the variation of QWS spacing $\Delta E$ near $E_F$ (within $[-1,1]$ eV) for different indexes $n$ (Figures 5c and 5d). The average level spacing $\overline{\Delta E}$ of QWSs for 61-ML Pb film is 222.5 (230.9) meV with (without) SOC, which is highly comparable with that (226.8 meV) in experiment (e.g., Figure 2d).

To explain the discreteness of $\Delta E$, the standard deviation $\sigma$ of $\Delta E$ has been calculated within different energy windows (insets of Figures 5c and 5d). Clearly, $\sigma$ with SOC within $[-0.5,0.5]$ and $[-1,1]$ eV is negligibly small. The ultralow $\sigma$ indicates $\Delta E$ for all QWSs is nearly unchanged, highlighting the relativistic-electron characteristics. As the energy window increases to $[-2,2]$ eV, the enhancement of $\sigma$ is still sufficiently small compared with $\Delta E$. This suggests the relativistic nature away from the band edge is the intrinsic property of Pb(111) films, similar to the observation for silicene on Ag(111).[53]

The following points may be noted:

i) The revealed linear $E(k_\perp)$ at $[-1,1]$ eV here falls within the reported energy windows for the quasi-linear $\Gamma L$ band in Pb as predicted by band-structure calculations (~$[-3,3]$ eV;[54] ~$[-2,4]$ eV[24, 25]). Different from the linear dispersions occurring near the conic Dirac point for topological materials, the nearly linear part of $\Gamma L$ band for Pb(111) is several eV near Fermi energy $E_F$ and far from the modified-parabolic-band edge at −4 eV (see right panel of Figure 5a). The $\Gamma L$ linearity only near $E_F$ is further supported by the sharp contrast of $\Delta E$ variation between regions near $E_F$ ($\Delta E$: homogeneous) and near band edge ($\Delta E$: inhomogeneous) (see left panel of Figure 5a). The relativistic-like physics indicated by the linear behavior of $E(k_\perp)$ also reconciles the high $v_F$ of $1.9\times10^6$ m/s,[4, 49] which is even larger than $v_F \approx 10^6$ m/s for graphene.[55] Physically, the surface geometric structure (triangular) of Pb(111) may induce the relativistic-type band near $E_F$, which is then further strengthened by SOC.

ii) With more degrees of freedom in $\mathbf{k}$ space, the 2D and 3D relativistic Dirac fermions in 3D topological insulators and Dirac semimetals show a cone structure with the spin-helical texture due to spin–momentum locking.[56] Differently, the linear $E(k_\perp)$ for Pb(111) is 1D, which obeys the Dirac equation and shares similarity in relativistic physics only along $k_\perp$ direction. Because of the 1D nature of linear $E(k_\perp)$ and the un-reported presence of Dirac point, Pb cannot be simply classified as topological insulator or Dirac semimetal.

iii) While QWS electrons extend over an energy scale of eV, Cooper-paired electrons condensate only within 1–2 meV near $E_F$. Accordingly, the QWSs and the derived relativistic fermions exist independent of whether Pb is superconducting.[4]

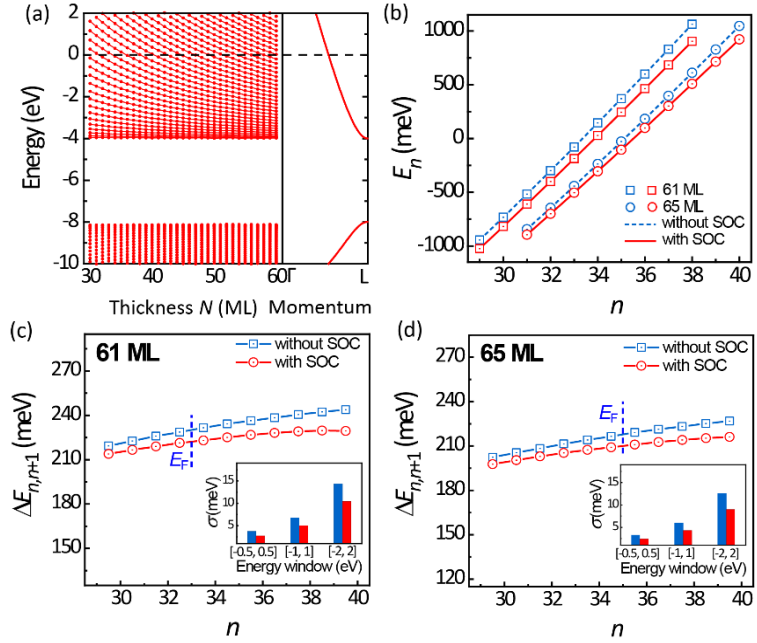

**Figure 5.** (a) Left panel: calculated QWSs energies at $\Gamma$ point as a function of thickness $N$. Right panel: bulk energy dispersion along $\Gamma L$ direction. The SOC is considered and the Fermi level is set at 0 eV. (b) Calculated energy positions $E_n$ of QWSs vs. indexes $n$ for 61- and 65-ML Pb(111) films with and without SOC. (c,d) Calculated energy-level spacing $\Delta E_{n,n+1}$ of QWSs for 61- and 65-ML Pb(111) with and without SOC. The insets are the standard deviation $\sigma$ of energy-level spacing $\Delta E_{n,n+1}$ within the energy windows of $[-0.5,0.5]$, $[-1,1]$, and $[-2,2]$ eV.

To the best of our knowledge, for intrinsic electronic systems, Pb(111) is the only platform beyond graphene showing strictly equally spaced quantum states besides Landau levels, implying the fundamental interest of confinement effect for relativistic particles in elementary quantum mechanics. The isolation of quasi-linear $\Gamma$–$L$ band from other calculated bands[24, 25, 54] 'protects' the relativistic QWSs by preventing the mixing with nearly-free-electron QWSs. Far from the $sp$-band edge, other metals, e.g., Ag,[33, 34] Au,[57] and Cu,[57] similarly show quasi-linearly dispersive band $E(k)$. The Pb(111) nano-island thus represents a class of metallic quantum-well structures of general interest for studying the relativistic-electron confinement.

Note also that exploring QWSs of the $\mathbf{k}_\perp$ Dirac fermions is unlikely in principle in topological insulators, and QWSs of 3D Dirac/Weyl fermions are rarely studied in nano-scale topological semimetals. Our results, together with all these considerations, may stimulate intensive investigations on the confinement of $k_∥$ Dirac quasiparticles in topological insulators and of 3D Dirac/Weyl quasiparticles in nano-structured topological semimetals. In applications, the analogy between relativistic electron and photon may further inspire developments of electron-type Fabry–Pérot interferometer, which can be analogously applied in binomial filter, wavemeter, and resonator, etc. Particularly, the unique behavior of equally spaced QWSs in Pb nano-islands may yield the sequence of same-frequency photon emissions repeating across different sets of adjacent QWSs, thereby creating a 'cascade' of emissions expected for the quantum cascade laser[9] in a simpler well structure.

■ **ASSOCIATED CONTENT**

**Supporting Information**

>Methods; more STM images of Pb nano-islands grown on epitaxial BLG/SiC; independent Fourier analyses of QWS spectra; additional examples of QWS modulations; QWS-spectra-derived information; discussion about BLG/SiC as the charge reservoir of relativistic electrons; Fabry–Pérot fittings to QWS spectra by incorporating parabolic $E(k_⊥)$ relation; discussion about negligible image potential in Pb nano-islands

■ **AUTHOR INFORMATION**

**Author contributions**

J.W. conceived and instructed the research. C.L. prepared the samples and carried out the STM experiments with assistance from C.C. C.Z., Z.Z., and Y.J. provided the first-principles calculations. C.L., C.Z., and S.Z. analyzed the data. C.L. and C.Z. wrote the manuscript with revisions from Z.Z., Y.J., and J.W.

**Notes**

The authors declare no competing financial interest.

■ **ACKNOWLEDGMENTS**

The authors acknowledge fruitful discussions with Feng Liu, Yan-Feng Zhang, Shuai-Hua Ji, Yu Li, Yi Liu, and Jun Ge. This work was financially supported by National Key R&D Program of China (No. 2017YFA0303302, and No. 2018YFA0305604), National Natural Science Foundation of China (No.11888101, No. 11774008, and No. 11774078), Beijing Natural Science Foundation (No. Z180010), and Strategic Priority Research Program of Chinese Academy of Sciences (No. XDB28000000).

■ **REFERENCES**

# Equally Spaced Quantum States in van der Waals Epitaxy-Grown Nano-Islands

Chaofei Liu,[†] Chunxiang Zhao,[‡] Shan Zhong,[†] Cheng Chen,[†] Zhenyu Zhang,[§] Yu Jia,[‡] and Jian Wang[†,#,¶,*]

[†]International Center for Quantum Materials, School of Physics, Peking University, Beijing 100871, China
[‡]International Laboratory for Quantum Functional Materials of Henan, and School of Physics and Microelectronics, Zhengzhou University, Zhengzhou 450001, China
[§]International Center for Quantum Design of Functional Materials (ICQD), Hefei National Laboratory for Physical Sciences at Microscale, and Synergetic Innovation Center of Quantum Information and Quantum Physics, University of Science and Technology of China, Hefei 230026, China
[#]CAS Center for Excellence in Topological Quantum Computation, University of Chinese Academy of Sciences, Beijing 100190, China
[¶]Beijing Academy of Quantum Information Sciences, Beijing 100193, China

[*]jianwangphysics@pku.edu.cn.

## Methods

*Sample growth and STM experiments.*—Our experiments were performed in an ultrahigh-vacuum ($5\times10^{-11}$–$2\times10^{-10}$ mbar) MBE–STM combined system. The graphitized 6*H*-SiC(0001) was prepared by the well-established cycles of flash annealing at 1300 °C.[1] The Pb nanoislands were grown with a deposition rate of 0.17 ML/min on BLG at 140–143 K by evaporating Pb from a standard Knudsen cell. At coverage of 3–7 ML, the islands normally show lateral size of the order of 50–100 nm. STM measurements, including topographic images and tunneling spectra (d*I*/d*V* vs. *V*), were conducted with a bias voltage equivalently applied to the sample at 4.25 K unless specified. The tunneling spectra were acquired using the standard lock-in technique with a bias modulation at 1.7699 kHz.

*Phenomenological simulations.*—In the Pb nanoislands/BLG system, the boundaries for the confined electrons originate from the island–vacuum and island–BLG interfaces. In a simple 1D model of an infinitely deep square-potential well, the fact that the electrons exist only in the form of standing waves dictates that $w = n\frac{\lambda}{2}$ (*w*, well width), which gives $k_\perp = \frac{n\pi}{w}$. Accordingly, the energy levels of the nonrelativistic electrons are quantized at $E_n = \frac{(\hbar k_\perp)^2}{2m} = \frac{(n\pi\hbar)^2}{2mw^2}$.[2] If we alternatively consider the relativistic massless electron with energy $E = \hbar k_\perp v_\text{F}$ ($v_\text{F}$, Fermi velocity) in the 1D infinite-well model, the quantized energy level follows $E_n = \frac{nhv_\text{F}}{2w}$, satisfying $E_n \propto n$ exactly as observed in our experiments [e.g., Fig. 2(d)].

The energy levels quantized at $E_n = \frac{nhv_\text{F}}{2w}$ for the relativistic fermions serve as the starting point of the phenomenological model. Due to the finite lifetime, the level at $E_n$ is broadened with finite width δ$_E$. A Lorentzian distribution is assumed to describe the broadened spectral lineshape,[3]

$$\rho(E - E_n) \propto \frac{1/\delta_E}{1+(E-E_n)^2(1/\delta_E)^2},$$

where *ρ* is the DOS. The DOS spectrum of QWSs is written as

$$\rho(E) = \sum_n \rho(E - E_n).$$

For the stacked well structure with different thicknesses, $w_1 = N_1 d$, $w_2 = N_2 d$ (*d*, interlayer spacing), the QWS spectrum is given by superimposing the energy spectra for different *N*,[3]

$$\rho(E) = \sum_N \sum_n \rho(E - E_{n,N}),$$

where $\rho(E - E_{n,N}) \propto \frac{1/\delta_E}{1+(E-E_{n,N})^2(1/\delta_E)^2}$. The spectral amplitude of the QWSs is renormalized by the piecewise exponential-decay function.

*Fabry–Pérot fittings.*—The QWS spectra can be analyzed based on the usual Fabry–Pérot formula, $I \propto \frac{1}{1+\frac{4f^2}{\pi^2}\sin^2\left(k_\perp N d + \frac{\Phi}{2}\right)} A(E)$.[4] Here, $f = \frac{\pi\sqrt{R}e^{-Nd/2l}}{1-Re^{-Nd/l}}$ (*R*, product of reflectivities at two boundaries; *N*, atomic layer number; *d*, interlayer spacing; $l$, mean free path), $\Phi$ is the total electron-scattering phase shift at surface and interface, and $A(E)$ is a smooth function for modulating the oscillation amplitude. The peak width $\Delta_E = \Gamma\eta \frac{1-Re^{-1/\eta}}{\sqrt{R}e^{-1/2\eta}}$, where $\Gamma$ is the quasiparticle inverse lifetime, and $\eta = \frac{l}{Nd}$. Normally, $\Delta_E > \Gamma$, except for $R = 1$ (ideal quantum well) and $l \gg Nd$. The quantities, $k_\perp$, $\Gamma$, $R$ and $\Phi$, completely specify the interferometer properties, which are thus of basic importance to solid-state physics. Specifically, $k_\perp$ and $\Phi$ give the QWS positions via Bohr–Sommerfeld quantization rule, $2k_\perp Nd + \Phi = 2n\pi$, and $\Gamma$ and $R$ give the QWS peak width via $\Delta_E = \Gamma\eta \frac{1-Re^{-1/\eta}}{\sqrt{R}e^{-1/2\eta}}$. Following Ref. [4], $\Gamma$, $R$ and $\Phi$ are parameterized as the quadratic functions of $E$.

*First-principles calculations.*—The QWS calculations of Pb films with different thicknesses were carried out using the density-functional theory (DFT) as implemented in the Vienna *ab initio* simulation package (VASP) with the projector augmented-wave method.[5-7] The generalized gradient approximation (GGA) developed by Perdew, Burke, and Ernzerhof (PBE)[8] was adopted for the exchange-correlation potential calculation. The 1×1 unit cells were used and the vacuum region of 16 atomic layers was adopted, which is thick enough for the system to converge to a correct total energy. The in-plane lattice parameter of the Pb(111) slabs is restricted to the experimental bulk value (4.95 Å). The Brillouin zone is sampled by 15×15×1 $k$-point meshes and the plane-wave energy cutoff is 500 eV. The convergence thresholds employed for energy and force are set to be $10^{-5}$ eV and $10^{-3}$ eV/Å, respectively. The Pb(111) films with the thickness of 30–65 ML were chosen in order to compare with our experimental results directly.

**Supplementary Text**

**I. Pb Nanoislands Grown on Epitaxial BLG/SiC**

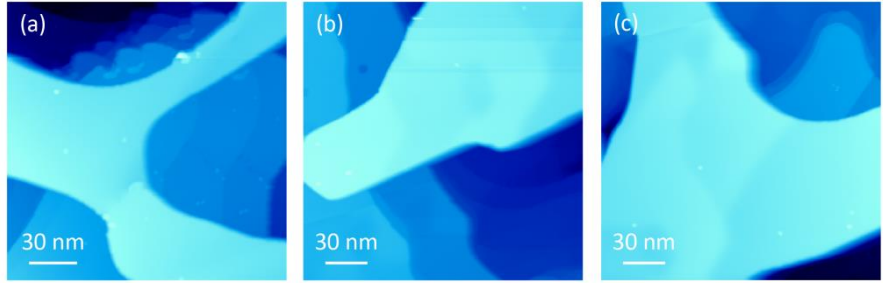

FIG. S1. Selected large-scale topographic images of the Pb nanoislands on BLG. Nominal coverage, 6.7 ML. The stacked irregular shapes are shown. Size: (a) 175×175 nm$^2$, (b,c) 200×200 nm$^2$; set point: (a–c) $V = 1$ V, $I = 500$ pA.

**II. More Fourier Analyses of QWS Spectra**

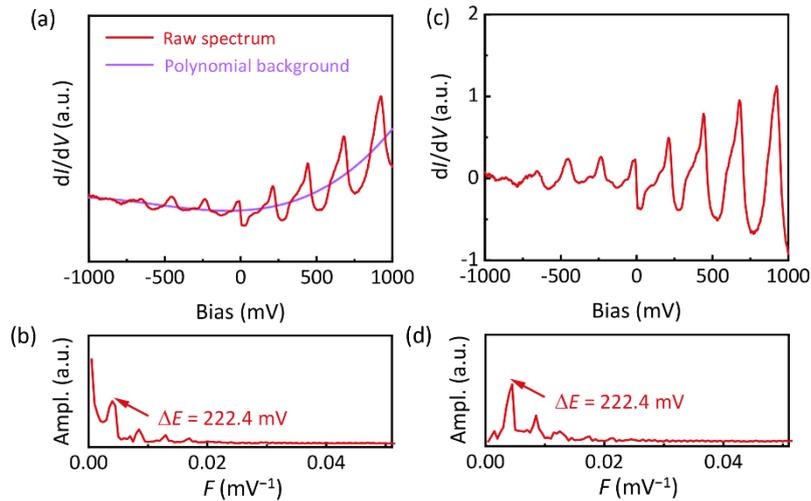

FIG. S2. Fourier analyses of raw and background-subtracted QWS spectra. (b) Fourier analysis of (a) the raw QWS spectrum [adapted from Fig. 2(b)]. The low-frequency peak (arrow) in (b) gives $\Delta E = 222.4$ meV. (c) QWS spectrum in (a) obtained via subtracting the polynomial background from its original spectrum. (d) Fourier analysis of the background-subtracted QWS

spectrum, giving $\Delta E$ = 222.4 meV.

## III. Uniformity of Tunneling Spectra

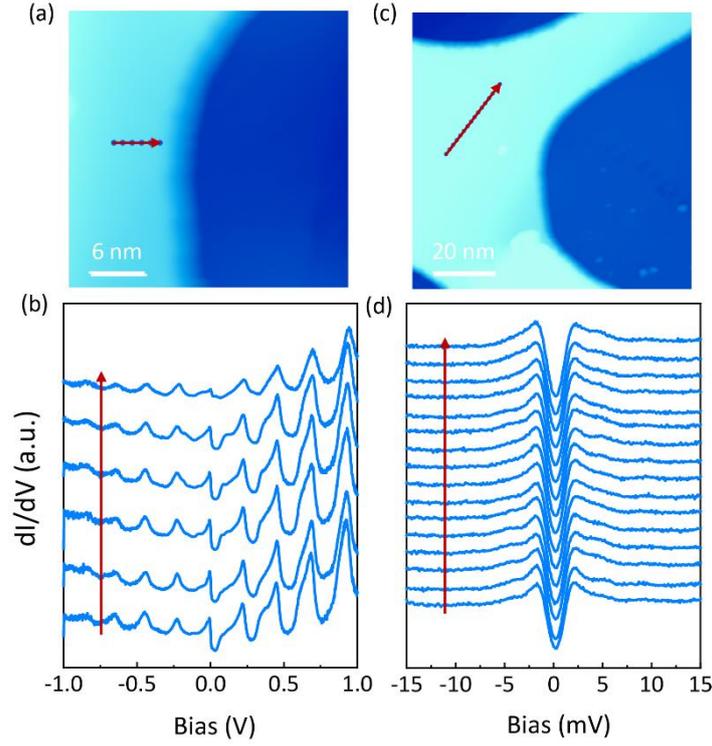

FIG. S3. (b,d) Tunneling spectra (vertically offset for clarity) taken on the positions in Pb nanoislands as indicated by arrows in (a,c) [adapted from Figs. 2(a) and 1(c)], respectively. (d) Set point: $V$ = 0.04 V, $I$ = 2500 pA; modulation: $V_{mod}$ = 0.3 mV.

## IV. Additional Examples of QWS Modulations

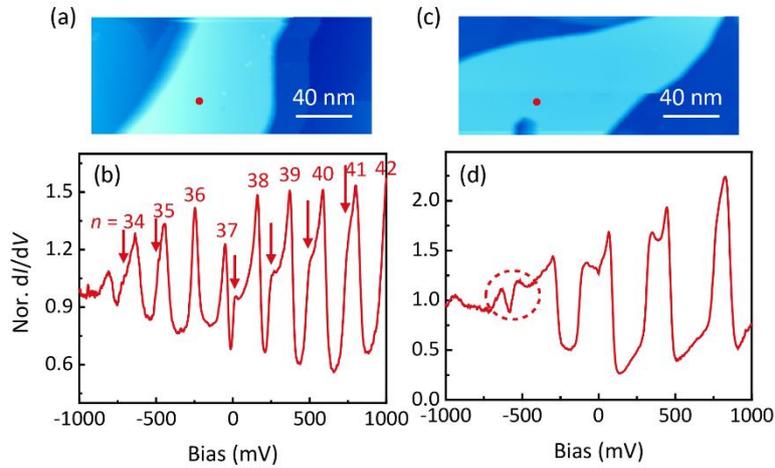

FIG. S4. Stacked QWSs. (b,d) Normalized QWS spectra taken on the positions in Pb nanoislands as indicated in (a,c) (nominal coverages, 6.7 ML), respectively. (a,c) Size: 200×85 nm$^2$; set point: $V$ = 1 V, $I$ = 500 pA.

The QWSs modulated by the discrete-structured Pb nanoislands exist universally (Fig. S4). In the modulated QWSs spectrum, the satellite peaks evolve periodically. In detail, for spectrum in Fig. S4(b), while $n$ = 42 QWS is a single peak, $n$ = 41–38 QWSs show side-peaks. As $n$ decreases from 41 to 38 in order, the side-peaks turn more obvious, and the intervals relative to their respective main QWSs gradually increase. Finally, at $n$ = 37, the side-peak 'disappears' and merges into the $n$ = 36 QWS, completing one evolution period. At $n$ = 35 and 34, the peaks of QWSs re-broaden, actually signaling the existence of satellite peaks, which might indicate next period of evolution. Nevertheless, due to the suppressed peak values of these two QWSs, the satellite–main peaks are difficult to be both resolved. For spectrum in Fig. 3(b), the periodic evolution of the satellite peaks is similar.

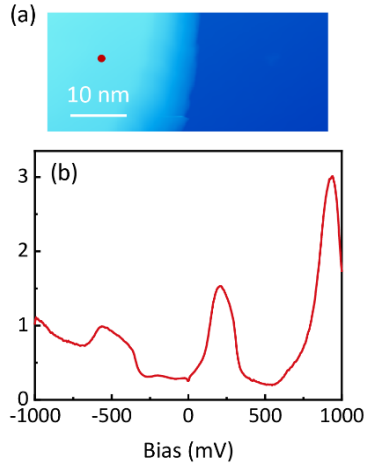

FIG. S5. QWSs with highly increasing period $\Delta E$, representing the typical spectrum reported previously. The spectrum was taken on the positions in Pb nano-island as indicated in (a) (nominal coverage, 3.3 ML; size: 50×21.25 nm$^2$; set point: $V$ = 0.5 V, $I$ = 500 pA). Theoretically, the interval $\Delta E$ of QWSs is anticorrelated with the well width.[9] The highly increased period $\Delta E$ is essentially due to the decreased island thickness as commonly expected for the QWSs.[10-22]

**V. QWS-Spectra-Derived Information**

The basic information derived from the QWS spectra is listed in Table S1. Because of the sensitivity to $w$, the detected QWSs can be used to distinguish $w$. Assuming Fermi velocity $v_F$ = 1.9×10$^6$ m/s along [111] direction as in Pb films,[13, 23] the linear fittings to the extracted $E_n$–$n$ relations for the Pb nanoislands by $E_n = \frac{nh v_F}{2w}$ around Fermi level (see Methods) yield $w$ as shown in Table S1. Strictly speaking, since the barriers for electrons confined in the islands are not ideally infinite, besides the island thickness, $w$ actually includes the effective thickness (~1 ML[14]) induced by the phase shift associated with reflections by the island–vacuum and island–substrate interfaces. Furthermore, the QWSs can strongly modulate the mechanical stability of the islands at different thicknesses.[24] Accordingly, expansion and shrinkage are exerted in the interlayer spacing, resulting in the thickness deviating from strict integer in unit of ML (2.86 Å along [111]).

In our experiments, the large-scale exposed graphitized SiC surface is difficult to obtain after the deposition of Pb nanoislands, highlighting the influence of Pb on the island–BLG/SiC interface. The well-defined Pb island–graphitized SiC boundary for directly measuring $w$ is only available for the situation of low-coverage Pb deposition [e.g., Fig. S5(a)]. Strikingly, for the Pb island in Fig. S5(a), the directly measured $w$ (5.2 nm) is highly comparable with that (5.25 nm; Table S1) determined from fitting the extracted $E_n$–$n$ data based on $E_n = \frac{nh v_F}{2w}$ ($\Delta E \equiv \frac{h v_F}{2w}$) for relativistic electrons, further proving the equally spaced QWSs as the evidence for confined relativistic-like fermions.

Table SI. Summary of QWS-spectra-derived information. $\Delta E$, interval for the main QWSs.

| QWS spectrum | $\Delta E$ (meV) | $w$ (nm) | [$N$] (ML) | [$n$] |
|---|---|---|---|---|
| Fig. 2 | 226.8 | 17.32 | 61 | 33 |
| Fig. 3(b) | 209.4 | 18.76 | 66 | 35 |
| Fig. 3(e) | 290.7 | 13.52 | 47 | 26 |
| Fig. S5(b) | 748 | 5.25 | 18 | 10 |

**VI. Discussion About BLG/SiC as the Charge Reservoir of Relativistic Electrons**

The interpretation of the equally spaced QWSs by the massless Dirac quasiparticles originating from the epitaxial BLG on SiC can be excluded.

i) The electronic structure $E(k)$ of BLG/SiC shows no linear behavior near $E_F$.

ii) BLG shows parabolic $E(k)$ near the Dirac point $E_D$ [Fig. 1(b): $E_D \in [-0.4,-0.3]$ eV],[25] yielding *non*-equally spaced

QWSs.

iii) Even though the fragmentary electronic structures for $E < E_D$ can be approximately taken as being linear, the associated energy window $[E, E_D]$ is generally limited to the order of $10^2$ meV. This is evidently far narrower than $[-1,1]$ eV, where the observed QWSs persistently exist with equal intervals.

## VII. Fabry–Pérot Fittings to QWS Spectra by Incorporating Parabolic $E(k_\perp)$ Relation

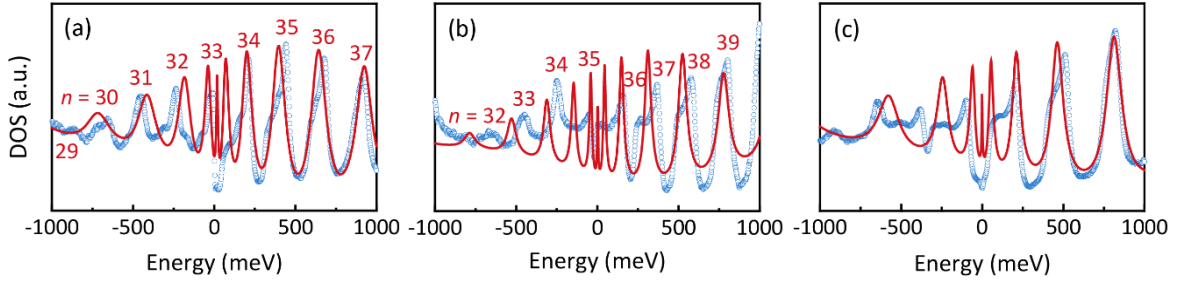

FIG. S6. Fittings to the main QWSs of 'normal' and 'stacked' spectra by Fabry–Pérot formula. The parabolic $E(k_\perp)$ relation (i.e., $k_\perp \propto \sqrt{E-c}$; $c$, fitting parameter) is incorporated and cannot fit the data.

## VIII. Negligible Image Potential in Pb Nanoislands

Notably, for ≲ 30-ML Pb film/Ag(111), despite the non-equally spaced QWSs within $[-1,4]$ eV, the extracted $E(k_\perp)$ still shows a quasi-linear dispersion,[10] seemingly contradicting our results. Actually, according to the Bohr–Sommerfeld quantization rule, $2k_\perp(E)Nd + \Phi(E) = 2n\pi$, the QWS interval is cooperatively set by $k_\perp(E)$ and $\Phi(E)$. At higher energies (≳1.2 eV), due to the collapse of a simple square-well scenario, the interval of QWSs is significantly modified by the image potential $\Phi_B(E)$ [component of $\Phi(E)$] near Pb surface.[26] By incorporating the phase accumulation of image potential, the anomalously shrinking interval of STM-measured QWSs ≳1.2 eV with increasing index $n$ can be theoretically reproduced.[26] In our experiments, the image-potential effect is negligible because of the following reasons.

i) The energy window for QWS spectra was set within $[-1,1]$ eV (< 1.2 eV) throughout to eliminate the image potential.

ii) Note that for Pb(111), $k_F d \approx \pi/2$ ($d$, interlayer spacing). The obtained $\Phi \lesssim \pi/2$ [Fig. 4(j)] gives a total electron penetration depth of ~$d/2$ into island–vacuum and island–BLG/SiC interfaces (notice that $\Phi$ includes phase contributions for electrons bouncing *back* and *forth*). The ~$d/2$ penetration is lower than previous report (~$d$),[13, 14] suggesting the absence of additional effective width introduced by the image potential $\Phi_B$.